\documentclass{webofc}
\usepackage[varg]{txfonts}   
\usepackage{upgreek}
\usepackage{bm}
\usepackage{tikz}

\usepackage{color,soul}

\setulcolor{red}

\newcommand{\bk}{{\bf k}}

\begin{document}
\title{Medium-modifications of $g\to c \bar{c}$ splitting 
}
%
%

\author{\firstname{Sohyun} \lastname{Park}\inst{1}\fnsep\thanks{\email{sohyun.park@cern.ch}} }

\institute{Theoretical Physics Department, CERN, 1211 Geneva 23, Switzerland
          }

\abstract{
We study medium-modifications of the gluon splitting into a quark and anti-quark pair. 
Applying the Baier-Dokshitzer-Mueller-Peigné-Schiff and Zakharov (BDMPS-Z) formalism, we derive a path-integral formula for the in-medium $g\to q \bar{q}$ splitting function in the close-to-eikonal limit. 
Our analysis shows that there are two qualitatively different medium effects: transverse momentum broadening of $q \bar{q}$ pairs and enhanced production of such pairs.  
We note that both effects are numerically sizeable if the average momentum transfer from the medium to the parton is at the quark mass scale.
In ultra-relativistic heavy-ion collisions, this condition is realized by charm quarks, therefore we focus our numerical analysis on the medium-modifications of $g\to c \bar{c}$ splitting.
}
\maketitle
\section{Introduction}
\label{intro}

Heavy quarks are produced by high-momentum transfer processes in hadronic collisions. 
In particular for a $c \bar{c}$ pair to be produced, 
the invariant mass $Q^2= \left( p_\text{c} + p_{\bar{\text{c}}} \right)^\mu \left( p_\text{c} + p_{\bar{\text{c}}} \right)_\mu$ should satisfy
$4 m_c^2 \leq Q^2 \leq \hat{s}$, where $m_c$ is the charm quark mass and $\sqrt{\hat{s}}$ is the partonic center of mass energy.
Since $m_c \gg \Lambda_{\rm QCD}$, this process is perturbatively calculable in QCD. 
Moreover, most $c \bar{c}$ pairs emerge back-to-back with $Q^2 \sim \mathcal{O}(\hat{s})$ at short distances, where the Quark Gluon Plasma (QGP) has not formed yet. 
That is, the total charm cross-section is mainly determined by such high-$Q^2$ short-distance processes and therefore almost unmodified by the QCD medium. 

In heavy-ion collisions, charm quarks produced at short distances enter the QGP, and the dominant in-medium process is gluon radiation off a charm, $c \to c\, g$ at long distances.
The BDMPS-Z calculation for the $c \to c\, g$ splitting function indicates that charm quarks traversing the QGP lose energy due to medium-induced gluon radiation~\cite{Baier:1996kr,Zakharov:1996fv,Dokshitzer:2001zm,Wiedemann:2000za,Gyulassy:2000er,Wang:2001ifa}.
This effect has been observed as the suppression of high-$p_T$ spectra of charmed hadrons. It should be noted however that this dominant process $c \to c\, g$ does not affect the total charm yield but only modifies the charm transverse momentum distribution.

The process we are focusing on is the medium-induced charm production with  $Q^2 \ll \hat{s}$ at long distances via a gluon splitting into a $c \bar{c}$ pair inside the QGP. 
This is much rarer compared to the charm production at short distances and sub-dominant compared to the medium-induced gluon radiation, but it is a dominant process for the in-medium charm production. 
We also remark two features making this process useful. First, the splitting rate depends on the degree of momentum transfer from the medium, hence it provides a way to probe the QGP properties. Second, the $c \bar{c}$ pair emerges collinear to the gluon direction, in contrast to back-to-back emergence in vacuum, so they are likely to remain inside the jet. Hence, splittings are traceable with jets.

\section{Medium-modified $g\to c\bar{c}$ splitting function}
\label{sec2_med-mod}

In the collinear limit $Q^2 \ll \hat{s}$, partonic cross-sections factorize. For example, the cross-section for the collision of two gluons producing a $c \bar{c}$ pair plus something else, say $X$, can be factorized by the cross-section for producing a parent gluon and the probability that this gluon splits into a $c \bar{c}$ pair,
\begin{equation}
	\hat{\sigma}^{g\, g \to c\, \bar{c}\, X} \,
	\xrightarrow{Q^2 \ll \hat{s}}\,  \hat{\sigma}^{g\, g \to g\, X} \, \frac{\alpha_s}{2\pi}\, \frac{1}{Q^2}\, P_{g \to c\bar{c}}(z) \,.	
	\label{eq1}
\end{equation}
Here the probability $P_{g \to c\bar{c}}(z)$ is called the $g\to c\bar{c}$ splitting function. In the collinear limit, the virtuality is given by 
$Q^2 = \tfrac{m_c^2 + \bm{\upkappa}^2}{z(1-z)}$, 
where 
$\bm{\upkappa} = \frac{1}{2}\left(\bk_c  - \bk_{\bar{c}} \right) $ is the relative momentum of the $c \bar{c}$ pair transverse to the gluon direction, and $z$ is the longitudinal momentum fraction carried by the charm. 
The in-vacuum $g\to c\bar{c}$ splitting function to leading order in $\alpha_s$ reads~\cite{Ellis:1996mzs}
\begin{equation}
\left(\frac{1}{Q^2}\, P_{g \to c\, \bar{c}} \right)^{\rm vac} (\bm{\upkappa}, z)
 =\frac{1}{2} \frac{1}{Q^4} \left[ \left( m_c^2 + \bm{\upkappa}^2 \right) \frac{z^2 + (1-z)^2}{z(1-z)}  + 2 m_c^2 \right]\, .
	\label{eq22.13}
\end{equation}
The in-medium $g\to c\bar{c}$ splitting function 
in the close-to-eikonal limit 
with the multiple soft scattering approximation
in the framework of the BDMPS-Z formalism  reads~\cite{Attems:2022ubu}
\begin{align}
&	 \left(\frac{1}{Q^2}\, P_{g \to c\, \bar{c}} \right)^{\rm med} (E_g, \bm{\upkappa}, z, \hat{q}, L)
		 		= 2\, \frak{Re}\, \frac{1}{8\, E_g^2}\, \int_{0}^{L} dt \int_t^{\infty} d\bar{t}\, 
    \exp\left[ i\frac{m_c^2}{2E_g z(1-z)} (t-\bar{t}) \right] \,
		\nonumber \\
		 &\qquad\qquad \times \int d{\bf r}_\text{out}
     \exp\left[ - \frac{1}{4} \int_{\bar{t}}^\infty d\xi\, 
     \hat{q}(\xi,z) \, {\bf r}^2_\text{out}  \right]\, 
		  \exp\left[{-i\, \bm{\upkappa} \cdot{\bf r}_\text{out}}\right]
		  \nonumber \\		  		
		&\qquad\qquad \times \left[ \left( m_\text{c}^2 + \frac{\partial}{\partial {\bf r}_\text{in}}\cdot \frac{\partial}{\partial {\bf r}_\text{out}}
		  \right) \frac{z^2 + (1-z)^2}{z(1-z)}  + 2 m_c^2  \right] \, {\cal K}\left[{\bf r}_\text{in}=0,t;{\bf r}_\text{out},\bar{t} \right] \,,
		    \label{eq22.2} 
\end{align}
where $E_g$ is the gluon energy and $L$ is the longitudinal size of the medium, and  
\begin{equation}
{\cal K}\big[{\bf x},t;{\bf r},\bar{t}\big] = \int_{{\bm{\uprho}}(t)}^{{\bm{\uprho}}(\bar{t})}
{\cal D}{\bm{\uprho}}\,  \exp\left[ i \int_t^{\bar t} d\xi\, \left(  \frac{E_g z(1-z)}{2}  \dot{\bm{\uprho}}^2 - \frac{\hat{q}(\xi,z)\,\bm{\uprho}^2}{4\, i} \right)\right] 	
\label{eq2.6}
\end{equation}
is the path-integral of a harmonic oscillator. 
This formula \eqref{eq22.2} describes the rate at which a gluon born at $t=0$ splits into a $c \bar{c}$ pair 
at longitudinal positions $t$ in amplitude and $\bar{t}$ in complex conjugate amplitude.
The medium effect is encoded in the so-called quenching parameter $\hat{q}(\xi,z)$ entering the absorption factor $\exp\left[ \cdots \right] $ in the second line and 
the path-integral ${\cal K}$ in the last line, which measures the squared momentum transferred per unit path-length from the medium to a parton.

Medium-induced shifts in transverse momentum arise in two occasions in the evolution,
one from back-propagating the phase $\exp\left[{-i\, \bm{\upkappa} \cdot{\bf r}_\text{out}}\right]$ from $\bar{t}$ to $\infty$ in the second line, and  
the other from acting the differential operator $\frac{\partial}{\partial {\bf r}_\text{in}}\cdot \frac{\partial}{\partial {\bf r}_\text{out}}$ on the path-integral 
in the last line of \eqref{eq22.2}.
Evaluating \eqref{eq22.2}, we find 
$
P_{g \to c\, \bar{c}} ^{\text{med}} \sim \mathcal{O}\left( \frac{\langle {\bf q}^2 \rangle_{\text{med}}}{Q^2} \right)
$.
We recall that the mass correction of $P_{g \to c\, \bar{c}} ^{\text{vac}}$ is of order $\mathcal{O}\left( \frac{m_c^2}{Q^2}\right)$. 
Therefore, medium-modifications become comparable to this mass correction at the scale 
\begin{equation}
\langle {\bf q}^2 \rangle_{\text{med}} = \int_{\tau_i}^{\tau_f} d\tau \hat{q}(\tau) \sim \mathcal{O}\left(m_c^2\right) \,.
\label{eq1.1}
\end{equation}
From model extraction in central PbPb collisions at $\sqrt{s_{NN}}=5.02\, \text{TeV}$, we obtain $4\, \text{GeV}^2 < \langle q^2\rangle_\text{med} < 8\, \text{GeV}^2$ \cite{Huss:2020whe}. That is, the condition \eqref{eq1.1} is realized for the charm mass of $m_c = 1.27\, \text{GeV}$. This is indeed why we focus on the $g \to c \bar{c}$ splitting although the formula \eqref{eq22.2} applies to gluons splitting into any quark-anti-quark pairs.

\section{Numerical results}
\label{sec3_numerical}

\begin{figure}[thb]
    \centering
    \includegraphics[width=0.40\textwidth,clip]{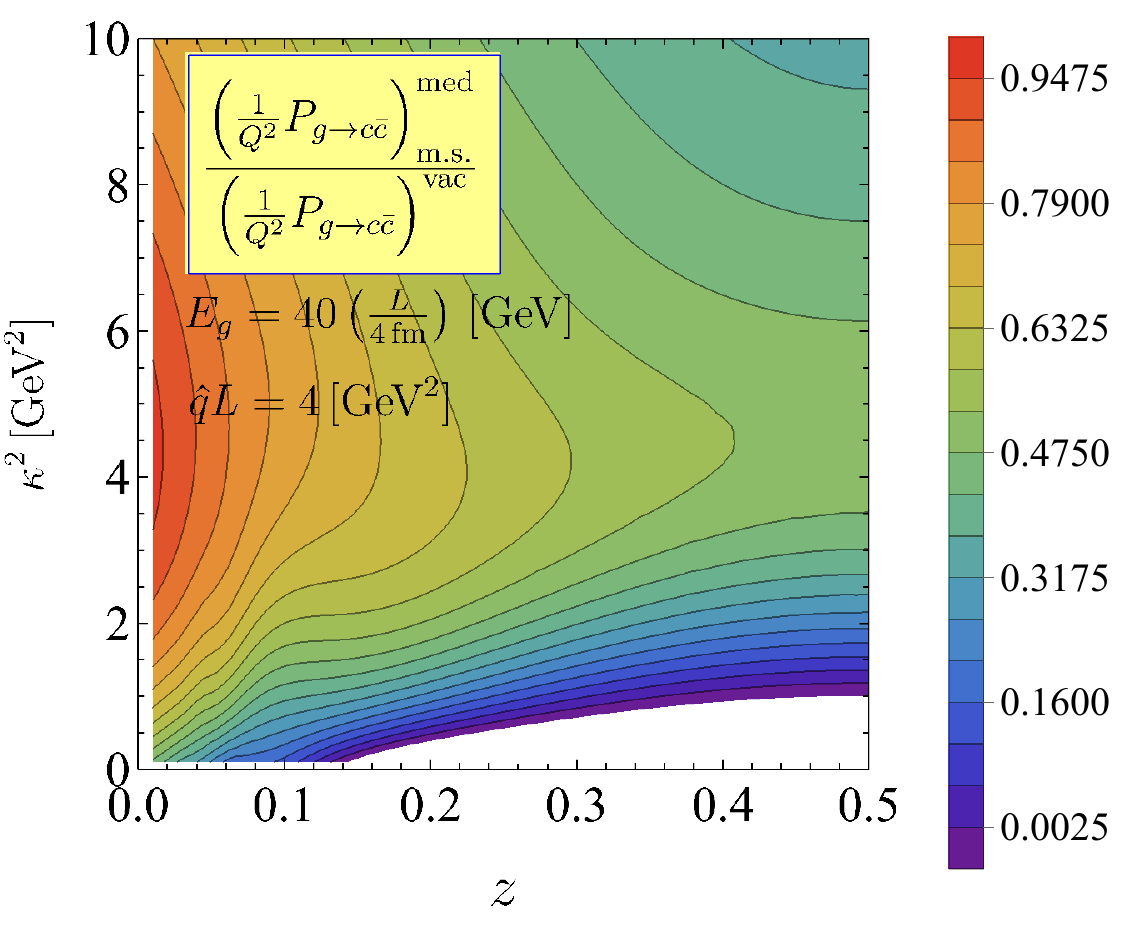}
    \hspace{0.03\textwidth}
    \includegraphics[width=0.53\textwidth,clip]{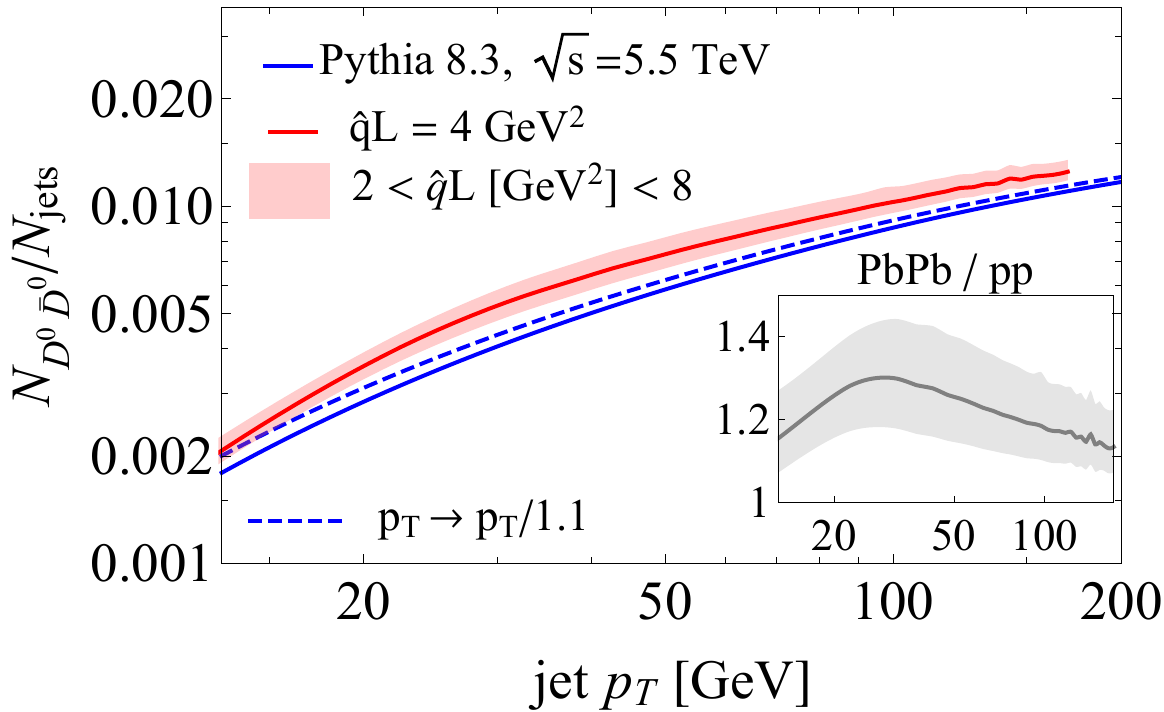}
    \caption{
    Left (from \cite{Attems:2022ubu}): 
    The ratio of the medium-modified splitting function to the vacuum one shows the characteristic momentum broadening and enhanced production of $c \bar{c}$ pairs. 
    Right (from \cite{Attems:2022otp}): 
    The fraction of jets carrying a $D^0\bar{D}^0$-tag in $\sqrt{s} = 5.5\, \text{TeV}$ mid-central pp collisions (blue line), and in central PbPb collisions (red band). The blue dashed line is included to take into account jet energy-loss estimated as $\Delta p_T/p_T \approx 10 \%$ of inclusive jets.
    }
    \label{fig-1}
\end{figure}

There are two signatures of medium-modification: 
transverse momentum broadening of $c\bar{c}$ pairs and enhanced production of such pairs.  
Left panel (from \cite{Attems:2022ubu}) of Figure~\ref{fig-1} plots numerical results of the medium-modification of \eqref{eq22.2} when the average squared momentum transferred from the medium is $\langle q^2\rangle_{\text{med}} = \hat{q}L = 4\, \text{GeV}^2$. It shows the depletion of low $\bm{\upkappa}^2$ splittings 
due to the in-medium broadening and enhanced production rate at the scale $\bm{\upkappa}^2 \sim \hat{q}\, L$. 
Qualitatively these two phenomena can be understood as follows: A gluon originally having a virtuality lower than the mass threshold $Q^2 < 4 m_c^2$ acquires transverse momentum as it traverses the medium, and hence the virtuality reaches or exceeds the threshold 
$Q^2 \geq 4 m_c^2$ so that the gluon can produce a $c \bar{c}$ pair, which results in the enhanced production. The charm and anti-charm produced also get momentum transfers in the medium, which leads to the momentum broadening.   

As an experimental strategy we propose to measure the ratio of the number of $D^0\bar{D}^0$-tagged gluon jets to the number of inclusive gluon jets as a function of jet $p_T$. Due to the enhanced production of $c \bar{c}$ pairs in the QGP, a larger fraction of $D^0\bar{D}^0$-tagged jets is expected in heavy-ion collisions than in pp collisions. 
Right panel~(from \cite{Attems:2022otp}) of Figure~\ref{fig-1} plots 
Monte Carlo simulation (using Pythia) results, which reveal $10 \% - 40 \%$ enhancement of $D^0\bar{D}^0$-tagged jets in heavy-ion collisions with respect to pp collisions~\cite{Attems:2022otp}. See the small box in the right corner. 
One remark is that for a very high $p_T$  correspondingly a very high gluon energy $E_g$, the enhancement is suppressed. 
This feature can be understood in terms of the formation time 
$
\tau_{\text{form}} 
$ 
of a $c\bar{c}$ pair. 
In the rest frame of the parent gluon, 
$
\tau_{\text{form, rest}} = \frac{1}{Q}.
$
In the lab frame, the gluon is boosted with respect to the medium by a Lorentz factor $\gamma = E_g/Q$, and hence dilated to 
$
\tau_{\text{form, boosted}} = \frac{E_g}{Q^2} \,.
$
If the gluon energy $E_g$ is too high, the formation time gets too delayed such that the splitting occurs after the gluon leaves the medium and hence no momentum transfers from the medium, i.e., no medium-induced production of  $c\bar{c}$ pairs.
  
Another remark concerns other mechanisms, besides the medium-modified $g \to c\bar{c}$ splitting, that can affect the ratio $N_{D^0\bar{D}^0}/N_\text{jets}$: 
First, jets lose energy to the QGP via the medium-induced gluon radiation $g\to gg$ and $q\to qg$, and as a result $p_T^\text{jet}$ decreases. We estimate the size of this jet energy-loss by the average fractional energy-loss $\Delta p_T/p_T \approx 10 \%$ of inclusive jets in central PbPb collisions~\cite{CMS:2011iwn} (blue dashed line in Figure~\ref{fig-1} Right).
Second, the medium-induced gluon radiation processes $g\to gg$ and $q\to qg$ change the distribution of gluons that can split into $c \bar{c}$ pairs and hence affect the rate of $c \bar{c}$ production inside jets. This effect certainly exists, however we have checked that it is numerically smaller and does not significantly affect our main conclusion of medium-enhanced $c \bar{c}$ production~\cite{Attems:2022otp}.

\section{Conclusion}
\label{sec4_conclusion}

We have calculated the medium-modification of the QCD leading order gluon splitting function into a charm and anti-charm pair in the BDMPS-Z formalism.
The result of  $P_{g \to c\bar{c}} ^{\text{med}}$ shows broadening of the relative transverse momentum of $c \bar{c}$ pairs and enhancement of $c \bar{c}$ production in the QGP, which are sizeable if the average momentum transfer from the medium is at the charm mass scale. 
As an experimental strategy for testing the enhanced $g \to c\bar{c}$ splittings, we have made a Monte Carlo study for the fraction of $D^0\bar{D}^0$-tagged jets over inclusive jets, which shows $10 \% - 40 \%$ enhancement for such measurements in heavy-ion collisions with respect to elementary collisions.

\section{Acknowledgment}
\label{sec5_Acknowledgment}
I thank Maximilian Attems, Jasmine Brewer, Gian Michele Innocenti, Aleksas Mazeliauskas, Wilke van der Schee and Urs Achim Wiedemann for collaboration and for critical readings of this manuscript. I also would like to thank D\'aniel Barta, Sangyong Jeon, Min-Jung Kweon, Saehanseul Oh and Jaebeom Park for useful questions and discussions during SQM 2022.

\end{document}